\documentclass[lettersize,journal]{IEEEtran}
\usepackage{amsmath,amsfonts}
\usepackage{algorithmic}
\usepackage{algorithm}
\usepackage{array}
\usepackage[caption=false,font=footnotesize,labelfont=rm,textfont=rm]{subfig}
\usepackage{textcomp}
\usepackage{stfloats}
\usepackage{url}
\usepackage{verbatim}
\usepackage{graphicx}
\usepackage{cite}

\usepackage{multirow}
\usepackage{booktabs}
\usepackage{xcolor}
\usepackage{hyperref}
\usepackage{pifont}
\usepackage{bm}

\begin{document}

\title{SQ-Whisper: Speaker-Querying based Whisper Model for Target-Speaker ASR}

\author{Pengcheng Guo, Xuankai Chang,~\IEEEmembership{Student Member,~IEEE}, Hang Lv, Shinji Watanabe,~\IEEEmembership{Fellow,~IEEE}, Lei Xie,~\IEEEmembership{Senior Member,~IEEE}
        % <-this % stops a space
}

\markboth{Journal of \LaTeX\ Class Files,~Vol.~14, No.~8, May~2024}%
{Shell \MakeLowercase{\textit{et al.}}: A Sample Article Using IEEEtran.cls for IEEE Journals}


\maketitle

\begin{abstract}
% Benefiting from massive and diverse data sources, speech foundation models demonstrate impressive performance across various tasks and are easily fine-tuned to new domains due to their knowledge transfer capabilities.
Benefiting from massive and diverse data sources, speech foundation models exhibit strong generalization and knowledge transfer capabilities to a wide range of downstream tasks.
However, a limitation arises from their exclusive handling of single-speaker speech input, making them ineffective in recognizing multi-speaker overlapped speech, a common occurrence in real-world scenarios.
% which is inevitable in real-world scenarios.
% However, adapting these models to recognize overlapped speech, such as target-speaker automatic speech recognition (ASR), is limited explored.
% In this paper, we present a study towards extending speech foundation models to target-speaker ASR (ASR).
In this study, we delve into the adaptation of speech foundation models to eliminate interfering speakers from overlapping speech and perform target-speaker automatic speech recognition (TS-ASR).
Initially, we utilize the Whisper model as the foundation for adaptation and conduct a thorough comparison of its integration with existing target-speaker adaptation techniques.
% Then, an innovative Speaker-Querying Transformer module is proposed to learn a sequence of speaker prompts from overlapping speech and target-speaker enrollment.
We then propose an innovative model termed Speaker-Querying Whisper (SQ-Whisper), which employs a set number of trainable queries to capture speaker prompts from overlapping speech based on target-speaker enrollment.
These prompts serve to steer the model in extracting speaker-specific features and accurately recognizing target-speaker transcriptions.
% We term this model the Speaker-Querying based Whisper (SQ-Whisper).
Experimental results demonstrate that our approach effectively adapts the pre-trained speech foundation model to TS-ASR.
Compared with the robust TS-HuBERT model, the proposed SQ-Whisper significantly improves performance, yielding up to 15\% and 10\% relative reductions in word error rates (WERs) on the Libri2Mix and WSJ0-2Mix datasets, respectively.
With data augmentation, we establish new state-of-the-art WERs of 14.6\% on the Libri2Mix Test set and 4.4\% on the WSJ0-2Mix Test set.
Furthermore, we evaluate our model on the real-world AMI meeting dataset, which shows consistent improvement over other adaptation methods.
% and achieves comparable results to the state-of-the-art model training with more overlapping data. 
% Our code is publicly available at~\href{https://github.com/pengchengguo/espnet/tree/tgtspk}{https://github.com/pengchengguo/espnet/tree/tgtspk}.
\end{abstract}

\begin{IEEEkeywords}
Speech Foundation Model, Whisper, Model Fine-tuning, Target-speaker ASR.  
\end{IEEEkeywords}

\section{Introduction} \label{sec:intro}
Large-scale foundation models have recently garnered immense interest due to their effectiveness across various deep learning tasks, such as natural language processing (NLP)~\cite{devlin2018bert, brown2020language, touvron2023llama2, achiam2023gpt, team2023gemini}, computer vision (CV)~\cite{dosovitskiy2020image, he2022masked, alayrac2022flamingo, li2023blip} and speech signal processing~\cite{mohamed2022self, radford2023robust, zhang2023google}.
% Trained on broad data at scale with billions of parameters, these models demonstrate strong generalization and knowledge transfer capabilities.
% Consequently, foundation models also serve as an excellent starting point for various downstream applications, excelling in zero-shot scenarios and being easily fine-tuned for new domains.
A notable strength of these models is their strong generalization ability on various tasks and domains, even in zero-shot scenarios. 
This ability arises from training on broad data at scale with billions of parameters, enabling them to capture a wide range of patterns and features.
Another significant advantage is their robust knowledge transfer capability, which makes them highly adaptable.
% to new tasks and domains.
By leveraging pre-trained knowledge, these general-purpose models can be tailored to new tasks through fine-tuning with small amounts of task-specific data, which significantly reduces the time and resources needed to train from scratch.
% This adaptability ensures that foundation models can quickly and effectively be tailored to meet specific requirements across different applications, enhancing their utility and impact in real-world scenarios.

In the realm of speech signal processing, various foundation models have been explored and shown promising performance.
% which can be broadly summarized into two categories depending on their training objective: self-supervised learning and supervised learning.
Depending on the training objective, they can be broadly summarized into two categories: self-supervised learning and supervised learning.
The former one aims to learn speech representations using audio-only data with contrastive loss~\cite{baevski2020wav2vec, conneau2020unsupervised, chung2021w2v}, generative loss~\cite{liu2020non, liu2021tera}, or predictive loss~\cite{hsu2021hubert, chen2022wavlm, baevski2022data2vec}.
Then, these models can be fine-tuned together with task-specific backbone networks or regarded as a feature extractor for downstream tasks, like automatic speech recognition (ASR), spoken language understanding (SLU), speaker identification, etc.~\cite{yang2021superb, chang2021exploration}.
As no label data is needed, this approach can be easily scaled to diverse speech data without annotation effort.
Supervised learning, on the other hand, aims to develop universal speech models capable of performing multiple speech tasks within a unified framework~\cite{zhang2023google, radford2023robust, peng2023reproducing, peng2024owsm}.
By jointly training on millions of multi-domain or multi-lingual labeled data, these models achieve high performance on different tasks, eliminating the need for data-specific fine-tuning.
% These models can directly be deployed in zero-shot scenarios or be fine-tuned with task-specific data for new domain.
% However, most of the existing foundation models only consider single-speaker speech input and may encounter challenges when confronted with multi-speaker overlapping scenarios, which is inevitable in real-world applications.

However, most existing foundation models only consider single-speaker speech input, lacking the ability to effectively prevent
interference from overlapping speech.
This capability is crucial in tackling the well-known cocktail party problem, in which multiple talkers speak simultaneously in noisy environments.
Constructing a foundation model for overlapping speech from scratch is impractical due to the extensive data collection and high training costs involved.
As a result, extending pre-trained foundation models to address this issue is of significant interest, and recent studies have made strides in this area.
For instance, Kanda \textit{et al.}~\cite{kanda2021large} pre-train a serialized output training (SOT) based multi-speaker ASR model with 900k hours of simulation data and then fine-tune it with a small amount of real meeting data.
Besides, Li \textit{et al.}~\cite{li2023adapting} develop an enhanced SOT method to adapt Whisper for multi-speaker ASR and diarization.
Apart from multi-speaker ASR, target-speaker ASR (TS-ASR), as one of the directions to recognize overlapping speech, also draws significant attention.
Methods such as target-speaker extraction (TSE)~\cite{huang2023adapting} and target-speaker HuBERT (TS-HuBERT)~\cite{zhang2023weakly} are based on self-supervised models, while prompt-tuned Whisper~\cite{ma2024extending} seeks to extend a supervised foundation model for TS-ASR.
% Similarly, in~\cite{ma2024extending}, Ma \textit{et al.} leverage prompt tuning to extend Whisper for target-speaker ASR.
Although these efforts have successfully transferred pre-trained foundation models to more complex scenarios, the results have not yet reached optimal performance, particularly for supervised foundation models~\cite{ma2024extending}.
This gap underscores the need for further research on supervised foundation models in the context of target-speaker adaptation.
% Therefore, in this work, we focus on adapting the supervised speech foundation model Whisper for TS-ASR.
%, one of the most famous supervised speech foundation models.

In this study, we aim to extend the capabilities of the existing speech foundation model, Whisper, to eliminate interfering speakers from overlapping speech and perform TS-ASR. 
Since the TSE~\cite{huang2023adapting} method has been successfully applied to self-supervised models, we first conduct a thorough exploration of its integration with the supervised speech foundation model and introduce TSE-Whisper.
% Specifically, we first conduct a thorough exploration of its integration with the target-speaker extraction (TSE)~\cite{li2023adapting} introduced for self-supervised pre-trained models.
We evaluate four types of TSE modules under both fully fine-tuning and parameter-efficient fine-tuning conditions.
Then, an innovative Speaker-Querying Whisper (SQ-Whisper) model is proposed, which employs trainable query parameters to learn 
a sequence of speaker prompts from overlapping speech given target-speaker enrollment.
These prompts are used to direct the Whisper model to extract speaker-specific features and accurately recognize target-speaker transcriptions.
Unlike TSE-based models, which rely on extra speaker models to extract speaker embeddings, our SQ-Whisper model dynamically captures target-speaker attributes for different overlapping scenarios and directly optimizes the speaker feature extraction using the ASR objective function.
Experimental results demonstrate the efficacy of our proposed method.
Compared with the robust TS-HuBERT~\cite{zhang2023weakly} model, the proposed SQ-Whisper achieves significant performance improvements, yielding up to 15\% and 10\% relative reductions in word error rates (WERs) on the Libri2Mix and WSJ0-2Mix simulation datasets, respectively.
With the inclusion of data augmentation, we establish new state-of-the-art WERs of 14.6\% on the Libri2Mix Test set and 4.4\% on the WSJ0-2Mix Test set.
% Furthermore, we also evaluate our model on the real-world AMI meeting dataset, which shows comparable results to the state-of-the-art model training with more overlapping data.
Furthermore, we evaluate our model on the real-world AMI meeting dataset, which shows consistent improvement over other adaptation methods trained with the same data size.
Note that our proposed method is a general framework and can be readily migrated to most Transformer based foundation models, such as HuBERT~\cite{hsu2021hubert} and OWSM series~\cite{peng2023reproducing, peng2024owsm}, similarly.
To facilitate research into SQ-Whisper's applications, we have made our recipes, setups, and code publicly available at~\href{https://github.com/pengchengguo/espnet/tree/tswhisper/egs2/librimix/tgt\_asr1}{https://github.com/pengchengguo/espnet/tree/tswhisper}.
% Our code is publicly available at~\href{https://github.com/pengchengguo/espnet/tree /tgtspk}{https://github.com/pengchengguo/espnet/tree/tgtspk}.

% We summarize our main contributions as follows:
% \begin{itemize}
%     \item We integrate the Whisper model with different established target-speaker adaptation methods and conduct a thorough comparison.
%     \item We propose an innovative SQ-Whisper model, which achieves superior performance on two multi-speaker overlapping simulation datasets and one real-world meeting dataset.
%     \item We make our code publicly available at at ~\href{https://github.com/pengchengguo/espnet/tree /tgtspk}{https://github.com/pengchengguo/espnet/tree/tgtspk}.
% \end{itemize}

% may encounter challenges when confronted with multi-speaker overlapping scenarios, which is inevitable in real-world applications

% Foundation models, also known as pre-trained models, are large-scale artificial intelligence (AI) models trained on vast amounts of data to acquire a deep understanding of language, images, or other modalities. These models serve as a starting point for various AI tasks, as they have already learned valuable data representations and can be fine-tuned or adapted for specific applications.

% We summarize our main contributions as follows:
% \begin{itemize}
%     \item We conduct a comprehensive exploration of integrating 
    
% \end{itemize}

\section{Related Work} \label{sec:related_work}

\subsection{Preliminary of the Whisper model} \label{subsec:detail_whisper}
% As one of the most famous speech foundation models, the OpenAI Whisper~\cite{radford2023robust} series has garnered tremendous attention for its incredible capabilities across various tasks, including language identification, multilingual speech recognition and translation, and utterance-level segmentation.
The Whisper model~\cite{radford2023robust} adopts an encoder-decoder architecture, with multiple Transformer blocks~\cite{vaswani2017attention} in both its audio encoder and text decoder.
% Given the input log-Mel filterbank features $\mathbf{X}$, the audio encoder first processes them through a 2-layer convolutional block (ConvBlock) for downsampling.
% Subsequently, the features are encoded into latent representations via multiple encoder blocks (EncoderBlocks).
Given the log Mel-filterbank input $\mathbf{X}$, the audio encoder initially processes these features through a 2-layer convolutional block (ConvBlock) for down-sampling, followed by multiple encoder blocks (EncoderBlocks) to generate latent representations.
% Subsequently, the features are encoded into latent representations via multiple encoder blocks (EncoderBlocks).
The process can be formulated as:
\begin{align}
    \mathbf{H}_{\text{conv}}  &= \text{ConvBlock} \left( \mathbf{X} \right)  \quad  \in \mathbb{R}^{T \times D_h}, \label{eq:cnn_out} \\
    \mathbf{H}_{\text{enc}}  &= \text{EncoderBlocks} \left( \mathbf{H}_{\text{conv}} \right) \quad \in \mathbb{R}^{T \times D_h}, \label{eq:enc_out}
\end{align}
where $T$ and $D_h$ represent the length and dimension of the hidden representations, respectively.
Subsequently, the text decoder predicts the probability distribution for the next token auto-regressively via stacked decoder blocks (DecoderBlocks):
\begin{equation}
    P\left( y_{l} \right) = \text{DecoderBlocks} \left( \mathbf{g}, \mathbf{y}_{1:l-1}; \mathbf{H}_{\text{enc}}\right), \label{eq:dec_out}
\end{equation}
where $y_{l}$ is the token to be predicted and $\mathbf{y}_{1:l-1}$ represents previous token sequence.
Here, $\mathbf{g}$ denotes a sequence of special tokens introduced by the Whisper model, such as the start token $\langle \left| \text{SOT} \right| \rangle$, end token $\langle \left| \text{EOT} \right| \rangle$, previous content token $\langle \left| \text{Prev} \right| \rangle$, language identifier $\langle \left| \text{En} \right| \rangle$, task identifier $\langle \left| \text{Transcribe} \right| \rangle$, etc.
The model is optimized to maximize the estimated probability of the target label with the standard cross-entropy (CE) loss:
\begin{equation}
    \mathcal{L}_{\text{CE}} = - \sum^{L}_{l=1} \log P(y_{l}), \label{eq:ce_loss}
\end{equation}
where $L$ means the length of transcription.

Trained on 680k hours of multi-lingual and multi-task weakly supervised data, the Whisper model exhibits remarkable capabilities across diverse tasks, including language identification, multilingual speech recognition and translation, and utterance-level segmentation.
% It even demonstrates strong generalization in zero-shot scenarios that are not included in its training set.
% Trained 680k hours of multi-lingual and multi-task weakly supervised data, the Whisper model demonstrates remarkable ASR capabilities and generalizes well in zero-shot scenarios.
% Furthermore, it also serves as a versatile starting point for fine-tuning towards new domains~\cite{gong2023whisper, wang2023whislu} or for distillation purposes in deployment~\cite{gandhi2023distil, shao2023whisper}.
Moreover, it also serves as a versatile starting point for fine-tuning towards new domains~\cite{gong2023whisper, wang2023whislu} or distillation in deployment scenarios~\cite{gandhi2023distil, shao2023whisper}.
% Besides, it can also be regarded as a starting point and be adapted to different downstream tasks via task-specific finetuning.
% However, it encounters difficulties and results in severe performance deteriorates when confronted with multi-speaker overlapping scenarios. This could be because the overlapping is not explicitly included in the Whisper multi-task training.

\subsection{Target-speaker ASR} \label{subsec:ts_asr}
% Although speech foundation models show remarkable speech recognition capabilities, they often encounter difficulties and result in severe performance deterioration when confronted with multi-speaker overlapping speech. 
While significant progress has been made in speech recognition, ASR systems often struggle with and experience severe performance degradation when confronted with multi-speaker overlapping speech.
Various studies have explored predicting the corresponding transcriptions for each speaker from overlapping speech, including permutation invariant training (PIT)~\cite{yu2017recognizing, chang2020end}, iterative prediction~\cite{guo2021multi}, serialized output training (SOT)~\cite{kanda2020serialized, liang2023ba} and their speaker-attributed variants~\cite{kanda2020joint, kanda2021comparative, li2023sa}.
However, these multi-speaker ASR methods may encounter limitations, such as predefined numbers of mixed speakers and challenges with permutation or speaker tracking during long audio inference~\cite{zhang2023conformer}.

In addition to multi-speaker ASR, target-speaker ASR (TS-ASR) represents a class of approaches for overlapping speech recognition.
TS-ASR aims to transcribe the desired speech of a target speaker from multi-speaker overlapping mixtures given the speaker's profile, like speaker embeddings or enrollment speech.
In~\cite{huang2023adapting}, Huang \textit{et al.} introduce a TSE module that adapts mixture representations through different speaker adaptation layers, enabling pre-trained self-supervised models in TS-ASR.
% In~\cite{huang2023adapting}, Huang \textit{et al.} introduce different speaker adaptation layers to enable pre-trained self-supervised models in target-speaker ASR.
In~\cite{zhang2023weakly}, Zhang \textit{et al.} propose TS-HuBERT, a self-supervised model that incorporates additional speaker enrollment to eliminate interfering speech and guides the learned representation towards the target speaker during pre-training.
Furthermore, Ma \textit{et al.}~\cite{ma2024extending} leverage prompt tuning to extend the Whisper model to TS-ASR.

% Target-speaker ASR, on the other hand, requires only one speaker profile of interest. It is apt for situations that require transcribing one target speaker while ignoring interfering speakers. By design, TS-ASR doesn't suffer from permutation ambiguity and speaker-tracing. However, it requires one inference per speaker if used to transcribe multiple speakers.

% \section{Proposed Model: TS-Whisper} \label{sec:ts_whisper}
% In this section, we delve into the adaptation of the pre-trained Whisper model to the target-speaker ASR task, denoted as target-speaker Whisper (TS-Whisper).
% Specifically, two types of models are explored: (1) the Baseline TS-Whisper model, which incorporates the speaker adaptation layers proposed in~\cite{huang2023adapting}, and (2) the Speaker-Querying TS-Whisper model, which utilizes an innovative Speaker-Querying Transformer module for adaptation.

% \subsection{Baseline TS-Whisper model} \label{subsec:baseline_ts_whisper}
% \section{Baseline TS-Whisper model} \label{sec:baseline_ts_whisper}

\begin{figure}[tbp]
    \centering
    \includegraphics[width=0.8\linewidth]{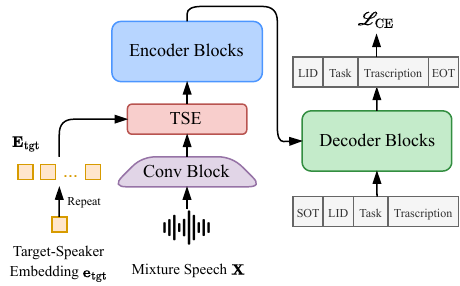}
    \caption{Overview of the TSE-Whisper model. ``TSE" is the target-speaker extraction module that extracts target-speaker features.}
    \label{fig:tse_whisper}
    % \vspace{-0.5cm}
\end{figure}

\section{Integrated: TSE-Whisper model} \label{sec:tse_whisper}
In~\cite{huang2023adapting}, the TSE module is inserted between the convolutional block and Transformer blocks of self-supervised models for joint fine-tuning.
It takes a pre-extracted speaker embedding vector $\mathbf{e}_{\text{tgt}} \in \mathbb{R}^{D_e}$ as an additional input and learns to extract only the target-speaker features from mixture representations, directing the model to predict target-speaker transcriptions.
Here, $D_e$ means the dimension of the speaker embedding.
We begin by integrating this TSE module into Whisper, defined as the TSE-Whisper model.
Fig.~\ref{fig:tse_whisper} provides an overview of the model.
The embedding vector $\mathbf{e}_{\text{tgt}}$ is first repeated $T$ times to get $\mathbf{E}_{\text{tgt}} \in \mathbb{R}^{T \times D_e}$, which has the same length as ConvBlock output $\mathbf{H}_{\text{conv}}$ (defined in Eq.~(\ref{eq:cnn_out})).
Then, four different TSE modules are evaluated to conduct target-speaker adaptation:

\begin{figure*}[tbp]
    \centering
    \includegraphics[width=0.8\textwidth]{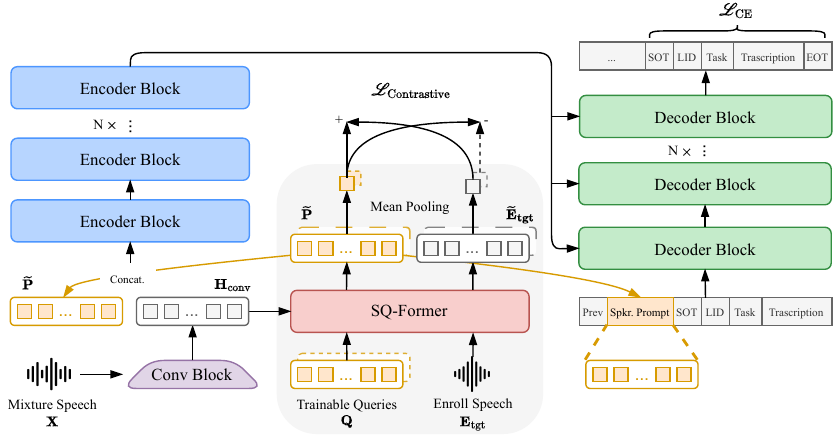}
    \caption{Overview of the proposed SQ-Whisper model. SQ-Former indicates the Speaker-Querying Transformer module for learning target-speaker prompts. $+$ and $-$ refer to positive and negative sample pairs, respectively, which are used to compute the speaker contrastive loss $\mathcal{L}_{\text{Contrastive}}$.}
    \label{fig:sq_whisper}
    % \vspace{-0.5cm}
\end{figure*}

\begin{enumerate}
    \item \textbf{Add}: $\mathbf{E}_{\text{tgt}}$ is directly added to $\mathbf{H}_{\text{conv}}$:
    \begin{equation}
        \mathbf{H}_{\text{tgt}} = \mathbf{H}_{\text{conv}} + w \left( \mathbf{E}_{\text{tgt}} \right) \quad \in \mathbb{R}^{T \times D_h},
    \end{equation}
    where $w: \mathbb{R}^{D_e} \to \mathbb{R}^{D_h}$ is a linear projection.

    \item  \textbf{Cat}: $\mathbf{E}_{\text{tgt}}$ and $\mathbf{H}_{\text{conv}}$ are concatenated together:
    \begin{equation}
        \mathbf{H}_{\text{tgt}} = w \left( \left[ \mathbf{H}_{\text{conv}}, \mathbf{E}_{\text{tgt}} \right]_f \right) \quad \in \mathbb{R}^{T \times D_h},
    \end{equation}
    where $\left[ \cdot, \cdot \right]_f$ denotes the concatenation operation along feature dimension and $w: \mathbb{R}^{D_h + D_e} \to \mathbb{R}^{D_h}$ is a linear projection.

    \item \textbf{FiLM}: The feature-wise linear modulation (FiLM)~\cite{perez2018film} layer carries out a feature-wise affine transformation conditioned on $\mathbf{E}_{\text{tgt}}$:
    \begin{equation}
        \mathbf{H}_{\text{tgt}} = w \left( \mathbf{E}_{\text{tgt}} \right) \circ \mathbf{H}_{\text{conv}} + b \left( \mathbf{E}_{\text{tgt}} \right) \quad \in \mathbb{R}^{T \times D_h},
    \end{equation}
    where $w: \mathbb{R}^{D_e} \to \mathbb{R}^{D_h}$ and $b: \mathbb{R}^{D_e} \to \mathbb{R}^{D_h}$ are linear projection layers, and $\circ$ is the element-wise product taken w.r.t the feature dimension.
    
    \item \textbf{CLN}: The conditional layer normalization (CLN)~\cite{pilault2020conditionally} estimates a speaker-specific scaling $\bm{\gamma}_{\text{tgt}}$ for the layer normalizations (LNs) in the first Transformer encoder block:
    \begin{align}
        \bm{\gamma}_{\text{tgt}} &= w \left( \mathbf{e}_{\text{tgt}} \right) \circ \bm{\gamma} + b \left( \mathbf{e}_{\text{tgt}} \right) \quad \in \mathbb{R}^{D_h}, \\
        \mathbf{H}_{\text{tgt}} &= \frac{ \mathbf{H}_{\text{conv}} - \bm{\mu}}{\bm{\sigma}} \circ r \left( \bm{\gamma}_{\text{tgt}} \right)+ \bm{\beta} \quad \in \mathbb{R}^{T \times D_h},
    \end{align}
    where $\bm{\gamma}$ and $\bm{\beta}$ are scaling and shifting factors in the standard LNs, while $\bm{\mu}$ and $\bm{\sigma}$ are the mean and standard deviation vectors of the input $\mathbf{H}_{\text{conv}}$.
    Here, $r: \mathbb{R}^{D_h} \to \mathbb{R}^{T \times D_h}$ is a repeat function.
    Essentially, $\bm{\gamma}_{\text{tgt}}$ is obtained by transforming $\bm{\gamma}$ with a FiLM layer conditioned on $\mathbf{e}_{\text{tgt}}$.
\end{enumerate}

The resulting target-speaker features $\mathbf{H}_{\text{tgt}}$ are then fed into the encoder blocks as illustrated in Eq.~(\ref{eq:enc_out}).

% \subsection{Speaker-Querying TS-Whisper model} \label{subsec:sq_ts_whisper}
% \section{Speaker-Querying based TS-Whisper model} \label{sec:sq_ts_whisper}
\section{Proposed: SQ-Whisper model} \label{sec:sq_whisper}

% In~\cite{li2023blip}, Li \textit{et al.} propose a lightweight Transformer module, called Querying Transformer (Q-Former), to bridge the modality gap between visual and textual inputs.
In~\cite{li2023blip}, Li \textit{et al.} propose a lightweight Transformer module, called Querying Transformer (Q-Former), to bridge the modality.
The Q-Former employs a set of learnable query vectors to interact with both the textual instructions and image encoder, extracting visual representations that are most informative of the text.
% The Q-Former employs a set of learnable query vectors to interact with both textual instructions and the image encoder through a two-stage training.
% In the first stage, it learns visual representations that are most informative of the text.
% In the second stage, it is trained to generate visual representations that can guide large language models (LLMs) in various downstream tasks, like image-to-text generation, image captioning, visual question answering, and more.
Inspired by it, we propose an innovative Speaker-Querying Transformer (SQ-Former) module, which aims to learn a sequence of speaker prompts conditioned on both multi-speaker mixture speech and target-speaker enrollment speech.
These prompts are then utilized to bootstrap the model to perform TS-ASR.
Unlike the two-stage pre-training of Q-Former, our SQ-Former functions as an adaptor and is jointly fine-tuned with the Whisper model. 
% We term this model the Speaker-Querying TS-Whisper model, and its overall framework is depicted in Fig.~\ref{fig:sq_ts_whisper}.
We term this resulting model the Speaker-Querying Whisper (SQ-Whisper).
Its overall framework is depicted in Fig.~\ref{fig:sq_whisper} and we will delve into the detailed discussion in the subsequent sections.

\begin{figure}[tbp]
    \centering
    \includegraphics[width=0.8\linewidth]{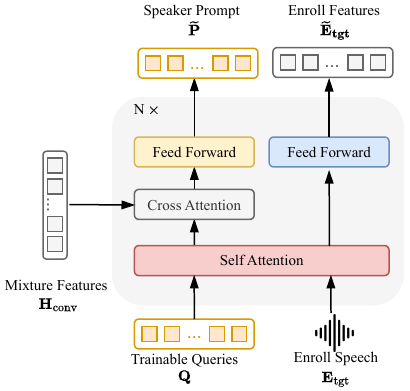}
    \caption{Details of the proposed SQ-Former module. Here, N$\times$ means to stack N blocks.}
    \label{fig:sq_former}
    % \vspace{-0.5cm}
\end{figure}

\subsection{SQ-Former module} \label{subsec:sq_former}
Fig.~\ref{fig:sq_former} illustrates the detailed architecture of the SQ-Former module.
It consists of multiple Transformer blocks and each block includes a self-attention (SelfAttn) module, a cross-attention (CrossAttn) module, and two distinct feedforward modules (FFNs) for the trainable queries and enrollment speech, respectively.

% Suppose we have multi-speaker mixture representations $\mathbf{H}_{\text{conv}}$ (as defined in Eq.~(\ref{eq:cnn_out})), a set of trainable query parameters $\mathbf{Q} \in \mathbb{R}^{L \times D}$, and the log Mel-filterbank features $\mathbf{E}_{\text{tgt}} \in \mathbb{R}^{T_e \times D_e}$ derived from the target-speaker enrollment speech.
% In this context, $L$ and $D$ represent the length and dimension of trainable queries, while $T_e$ and $D_e$ denote the length and dimension of enrollment features.
Suppose we have multi-speaker mixture representations $\mathbf{H}_{\text{conv}}$ (as defined in Eq.~(\ref{eq:cnn_out})) and the log Mel-filterbank features $\mathbf{E}_{\text{tgt}} \in \mathbb{R}^{T_e \times D_e}$ derived from the target-speaker enrollment speech.
In this context, $T_e$ and $D_e$ denote the length and dimension of enrollment features.
The SQ-Former module creates a set of trainable query vectors $\mathbf{Q} \in \mathbb{R}^{L_q \times D_q}$ as additional input to learn speaker prompts.
Here, $L_q$ and $D_q$ represent the length and dimension of trainable query vectors.
The computation of SQ-Former can be broken down into three steps:
\begin{enumerate}
    \item \textbf{Learn}: Initially, the trainable query $\mathbf{Q}$ created by SQ-Former interacts with $\mathbf{E}_{\text{tgt}}$ through a SelfAttn module to learn the characteristics of target speaker:
    \begin{equation}
        \left[ \mathbf{\tilde{Q}}, \mathbf{\tilde{E}}_{\text{tgt}} \right] = \text{SelfAttn} \left( \left[ \mathbf{Q}, w \left( \mathbf{E}_{\text{tgt}} \right) \right]_l \right) \in \mathbb{R}^{(L_q + T_e) \times D_q},
    \end{equation}
    where $w: \mathbb{R}^{D_e} \to \mathbb{R}^{D_q}$ is a linear projection and $\left[ \cdot, \cdot \right]_l$ denotes the concatenation operation along the length dimension.
    % Through the attention mechanism, the $\mathbf{\tilde{Q}}$ gathers the speaker's attributes.

    \item \textbf{Search}: Next, $\mathbf{\tilde{Q}}$ attends to $\mathbf{H}_{\text{conv}}$ through a CrossAttn module to search for features that most relevant to the target speaker:
    \begin{equation}
        \mathbf{ \tilde{P} } = \text{CrossAttn} \left( \mathbf{\tilde{Q}}, b \left( \mathbf{H}_{\text{conv}} \right) \right) \quad \in \mathbb{R}^{L_q \times D_q},
    \end{equation}
    where $b: \mathbb{R}^{D_h \to D_q}$ is a linear projection and $\mathbf{ \tilde{P} }$ denotes the obtained target-speaker features (i.e., speaker prompts). For the CrossAttn module, $\mathbf{\tilde{Q}}$ is regarded as the query vectors while $\mathbf{H}_{\text{conv}}$ is the key and value vectors.

    \item \textbf{Transform}: Two FFNs are used for transformation. Each FFN consists of two linear projections with a GeLU activation function in between. The computational process can be described as:
    \begin{align}
        \mathbf{\tilde{P}} &= \text{FFN}_1 \left( \mathbf{\tilde{P}} \right) \quad \in \mathbb{R}^{L_q \times D_q}, \label{eq:final_prompt} \\
        \mathbf{\tilde{E}_{\text{tgt}}} &= \text{FFN}_2 \left( \mathbf{ \tilde{E}_{\text{tgt}} } \right) \quad \in \mathbb{R}^{T_e \times D_q}. \label{eq:final_enroll}
    \end{align}
\end{enumerate}
Note that we omit the residual connection and layer normalization of the Transformer block for clarity.

The difference between the Q-Former and our SQ-Former lies in their motivation: the Q-Former acts as an information bottleneck to summarize visual features given text instructions, whereas the SQ-Former aims to retrieve the attributes of target speakers from multi-speaker mixture speech based on the search term, i.e. target-speaker enrollments.
Besides, we also introduce an additional speaker contrastive loss to help the SQ-Former generate more discriminative speaker prompts, which will be elaborated on in the following section.

\subsection{Speaker contrastive loss} \label{subsec:con_loss}
The speaker contrastive loss is adopted to identify the true speaker prompts within $K$ distractors of enrollment features.
In detail, the speaker prompts $\mathbf{\tilde{P}}$ and the enrollment features $\mathbf{\tilde{E}}_{\text{tgt}}$ (as in Eq.~(\ref{eq:final_prompt}) and Eq.~(\ref{eq:final_enroll})) first undergo mean pooling to yield global vectors $\mathbf{\bar{p}} \in \mathbb{R}^{D_q}$ and $\mathbf{\bar{e}_{\text{tgt}}} \in \mathbb{R}^{D_q}$, respectively.
Then, the speaker contrastive loss is calculated as follows:
\begin{equation} \label{eq:con_loss}
    \mathcal{L}_{\text{Contrastive}}  = - \log \frac{ \exp \left( \cos \left( \mathbf{\bar{p}}, \mathbf{\bar{e}_{\text{tgt}}}\right) / \kappa \right)} { \sum_{\mathbf{\bar{e}} \sim D\left( \mathbf{\bar{e}}_{\text{tgt}} \right)} \exp \left( sim \left( \mathbf{\bar{p}}, \mathbf{\bar{e}}\right) / \kappa \right)},
\end{equation}
where $\cos \left( \cdot \right)$ represents the cosine similarity between two vectors and $\kappa$ is a temperature hyper-parameter. 
Here, $D\left( \mathbf{\bar{e}}_{\text{tgt}} \right)$ is a set comprising all candidate enrollment features, which includes the paired $\mathbf{\bar{e}}_{\text{tgt}}$ as a positive sample and enrollments from other speakers in the same batch as negative samples.

The final loss function of our SQ-Whisper model is given by:
\begin{equation} \label{eq:final_loss}
    \mathcal{L} = \mathcal{L}_{\text{CE}} + \alpha \mathcal{L}_{\text{Contrastive}},
\end{equation}
where $\alpha$ refers to an interpolation factor that scales the speaker contrastive loss.
We sample $K=10$ negative samples during the training.
Rather than extensively fine-tuning $\alpha$, we experiment with values of 10, 20, and 30 to align the magnitudes of two loss functions.
We observe that varying $\alpha$ within this range has minimal impact on the performance; thus, we set $\alpha = 20$ for all experiments to maintain consistency. 

\subsection{Prompt scheme} \label{subsec:prompt_scheme}
% , derived from Eq.~(\ref{eq:final_prompt}),
The learned target-speaker prompts $\mathbf{\tilde{P}}$ are first transformed to maintain consistency with the size of hidden representations and then utilized in both the audio encoder and text decoder.
Specifically, we steer the encoder to extract target-speaker features by appending the prompts to the input features.
% \begin{equation}
%     \mathbf{H}_{\text{enc}} = \text{EncoderBlocks} \left( \left[ \mathbf{\tilde{P}}, \mathbf{H}_{\text{conv}} \right] \right) \in \mathbb{R}^{(L+T) \times D},
% \end{equation}
% where $\left[ \cdot, \cdot \right]$ denotes the concatenation operation along the length dimension, while $L$ and $T$ are the lengths of speaker prompts and speech features, respectively. 
For the decoder, the prompts are inserted between the embedding space of the special token $\langle \left| \text{Prev} \right| \rangle$ and token $\langle \left| \text{SOT} \right| \rangle$.
These prompt schemes are distinguished as ``Enc. Prompt" and ``Dec. Prompt", correspondingly.
Fig.~\ref{fig:sq_whisper} provides a schematic of the prompt scheme.
Note that the output labels corresponding to the speaker prompts are excluded from the computation of the CE loss.

\subsection{Parameter efficient fine-tuning} \label{subsec:peft}
Fully fine-tuning large-scale foundation models to new datasets or tasks can be challenging due to their extensive parameter size.
Therefore, we also explore the utilization of parameter-efficient fine-tuning (PEFT) methods when adapting the Whisper models.
In this study, we adopt the widely-used low-rank adaptation (LoRA)~\cite{hu2022lora} approach.
The main idea of LoRA is to inject trainable rank decomposition matrices into each layer of the Transformer model while freezing all well-trained weights.
Specifically, for each liner projection $w: \mathbb{R}^{D_\text{in}} \to \mathbb{R}^{D_\text{out}}$ in a Transformer layer, where $D_\text{in}$ and $D_\text{out}$ represent the input and output dimensions, two new projections $w_a: \mathbb{R}^{D_\text{in}} \to \mathbb{R}^{R}$ and $w_b: \mathbb{R}^{R} \to \mathbb{R}^{D_\text{out}}$ are introduced, where $R$ is the rank number and $R \ll \operatorname{min}(D_\text{in}, D_\text{out})$.
The modified forward process can be formulated as:
\begin{equation} \label{eq:lora}
    \mathbf{h_\text{out}} = w \left( \mathbf{h_\text{in}} \right) + w_b \left( w_a \left( \mathbf{h_\text{in}} \right) \right),
\end{equation}
where $\mathbf{h_\text{in}}$ and $\mathbf{h_\text{out}}$ means the input and output hidden representations.
During LoRA fine-tuning, only the parameters of $w_a$ and $w_b$ are updated, resulting in a significant reduction of memory footprint.

% Specifically, for each weight $\mathbf{W} \in \mathbb{R}^{D_1 \times D_2}$ in a Transformer layer, where $D_1$ and $D_2$ represent the input and output dimensions, two new matrices $\mathbf{W}_a \in \mathbb{R}^{D_1 \times R}$ and $\mathbf{W}_b \in \mathbb{R}^{R \times D_2}$ are introduced, where $R$ is the rank number and $R \ll \operatorname{min}(D_1, D_2)$.
% During fine-tuning, the input features are multiplied with both the original weight $\mathbf{W}$ and the low-rank matrices $\mathbf{W}_a$ and $\mathbf{W}_b$.
% Then the two outputs are summed and fed into the next layer.
% It is important to note that only $\mathbf{W}_a$ and $\mathbf{W}_b$ are updated, while $\mathbf{W}$ is frozen, resulting in a significant reduction in memory footprint.
% The forward process becomes:

% \begin{equation} \label{eq:lora}
%     % \mathbf{x} = \mathbf{W}\mathbf{x} + \triangledown\mathbf{W}\mathbf{x} = \mathbf{W}\mathbf{x} + \mathbf{W}_b\mathbf{W}_a\mathbf{x}
%     \mathbf{H} = \mathbf{W}\mathbf{H} + \mathbf{W}_{b}\mathbf{W}_{a}\mathbf{H}.
% \end{equation}

\section{Experimental Setups} \label{sec:exp_setup}
In this section, we will describe the datasets, detail the setups of the TSE-Whisper and SQ-Whisper models, and present the hyper-parameters used for training and inference.

\subsection{Datasets} \label{subsec:data}
Our proposed model is evaluated on two commonly used mixture simulation datasets: Libri2Mix~\cite{cosentino2020librimix} and WSJ0-2Mix~\cite{hershey2016deep}, and a real-world meeting dataset: AMI~\cite{carletta2005ami}.

The Libri2Mix dataset is created by randomly mixing multiple source utterances from different speakers in the LibriSpeech~\cite{panayotov2015librispeech} along with noise samples from the WHAM!~\cite{wichern2019wham}.
It comprises both clean and noisy conditions, with each including two training sets (Train-$100$, Train-$360$), one validation set (Dev), and one test set (Test).
The signal-to-noise ratios (SNRs) for the clean mixtures follow a normal distribution with a mean of $0$ dB and a standard deviation of $4.1$ dB, while for the noisy mixtures, they are $-2$ dB and $3.6$ dB.
Our experiments are primarily conducted on the noisy Train-$100$ subset, unless otherwise specified.
The Train-$100$, Dev, and Test sets contain 13900, 3000, and 3000 mixture samples, respectively.
During the training, the target-speaker enrollment is randomly selected from other training utterances of the same speaker, while for the inference phase, we utilize the existing enrollment lists provided in the SpeakerBeam codebase\footnote{\url{https://github.com/BUTSpeechFIT/speakerbeam}}.
% ~\cite{vzmolikova2019speakerbeam}

The WSJ0-2Mix dataset is generated by mixing pairs of utterances from different speakers at random SNRs, derived from the WSJ0~\cite{paul1992design}.
There are 4 variations of the dataset, representing 2 distinct sampling rates (16 kHz and 8 kHz) and 2 mixing modes (min and max).
In the min mode, the mixture stops with the shortest utterance, while the max mode pads the shortest utterance to match the length of the longest one.
We use the max\_$16$k sets for our experiments, which have 20000, 5000, and 3000 mixture samples for training, validation, and evaluation, respectively.
% Its training and validation sets share common speakers from the si\_tr\_s subset and the test set combines both si\_dt\_05 and si\_et\_05 subsets.
The selection for the target-speaker enrollment follows the same strategy as mentioned above, while the enrollment list is sourced from the SpEx\_Plus implementation\footnote{\url{https://github.com/gemengtju/SpEx_Plus}}.
% ~\cite{ge2020spex+}

The AMI dataset comprises approximately 100 hours of real-world English meeting recordings.
Each meeting involves 3 to 5 participants and is recorded using headset microphones and two 8-channel microphone arrays.
For our experiments, we utilize the single channel distant microphone (SDM1) audio set for experiments.
The overlap ratio of this set is around 12\%.
We segment each recording using the human-annotated timestamps, which means the audio length is based on the shorter speech and the interfering speech will be regarded as noise.
Since the same speaker appears in multiple meetings, their clean segments (IHM) from different meetings are selected as enrollment speech.

\subsection{Setups} \label{subsec:setups}
\subsubsection{TSE-Whisper model} \label{subsubsec:tse_whisper_setup}
We consider the TSE-Whisper model that integrates previous prominent TSE modules (described in Section~\ref{sec:tse_whisper}) as our baseline.
The Whisper model used for fine-tuning is the multilingual medium variant, which consists of 2 convolutional layers, 24 encoder blocks, and 24 decoder blocks.
Each encoder block has 16 attention heads with a total attention dimension of 1024 ($D_h = 1024$) and a feedforward dimension of 4096.
The decoder block mirrors the encoder's structure, except for an additional cross-attention layer featuring 16 attention heads and 1024 attention dimensions.
For the extraction of target-speaker embedding $\mathbf{e}_{\text{tgt}}$, we employ a ResNet-34 speaker verification model\footnote{\url{https://github.com/wenet-e2e/wespeaker/blob/master/docs/pretrained.md}}, pre-trained on the VoxCeleb~\cite{nagrani2017voxceleb} dataset, yielding 256-dimensional embeddings ($D_e = 256$).
% We utilize a ResNet-34 speaker verification model~\cite{wang2023wespeaker}\footnote{\url{https://github.com/wenet-e2e/wespeaker/blob/master/docs/pretrained.md}} pre-trained on the Voxceleb~\cite{nagrani2017voxceleb} dataset to extract 256-dimensional target speaker embeddings as enrollments.

\subsubsection{Proposed SQ-Whisper model} \label{subsubsec:sq_whisper_setup}
For the proposed SQ-Whisper, the same multilingual Whisper medium model is adopted as the foundation model.
%, unless stated otherwise.
The SQ-Former module, as detailed in Section~\ref{subsec:sq_former}, is built upon the $\text{BERT}_{\text{base}}$~\cite{devlin2018bert} encoder.
It consists of 2 blocks, each containing a self-attention layer, a cross-attention layer, and 2 distinct feedforward layers dedicated to the trainable queries and enrollment speech, respectively.
The number of attention heads is 12 and the dimension of hidden units is 768 ($D_q = 768$).
All SQ-Former weights are initialized randomly and trained from scratch.
We use 16 trainable query vectors ($L_q = 16$) with each having a dimension of 768 (same as the hidden units of the SQ-Former).
The number of queries is tuned from 1 to 64 in the ablation analysis of Section~\ref{subsec:num_queries_results}.

\subsubsection{Training configurations} \label{subsubsec:train_config}
All of the models are implemented using the ESPnet toolkit~\cite{watanabe2018espnet}.
Data preparation follows ESPnet recipes but without speech perturbation and SpecAugment for a fair comparison with~\cite{zhang2023weakly} if not specified.
Each model is trained for 10 epochs and the best 5 checkpoints are averaged for the final inference.
We use one 48GB A6000 GPU for all experiments.
A warm-up scheduler is employed to adjust the learning rate, peaking at $5e-5$ with 1500 steps.
We use the greedy search decoding strategy for the inference.
% In LoRA fine-tuning, we apply LoRA to four attention weights in each encoder and decoder block (e.g., $\textbf{W}_q$, $\textbf{W}_k$, $\textbf{W}_v$, $\textbf{W}_o$), with the LoRA rank ($R$) set to 16.
In LoRA fine-tuning, we apply LoRA to four attention weights in the original Whisper encoder and decoder (e.g., query, key, value, and output projection layers), with $R=16$.
It is important to note that both LoRA parameters and the added target speaker adapter are trainable during LoRA fine-tuning.
% The detailed configurations, implementation, and pre-trained models will be made available later.

\section{Experimental Results and Analysis} \label{sec:results_analysis}
In this section, we will compare our proposed model with various systems and perform ablation studies to investigate its effectiveness.

\subsection{Results of the TSE-Whisper model} \label{subsec:tse_whisper_results}

\begin{table}[tbp]
    \centering
    \caption{Word error rates (WERs \%) of our TSE-Whisper models on the noisy Libri2Mix sets. ``\#Param" refers to the number of trainable parameters. ``Full" means fully fine-tuning all parameters, while ``LoRA" means LoRA fine-tuning.}
    \resizebox{1.0\linewidth}{!}{
        \begin{tabular}{lcccc} 
            \toprule
            \multirow{2}{*}{\textbf{Model}} & \multirow{2}{*}{\textbf{Adapt. Method}} & \multirow{2}{*}{\textbf{ \#Param}} & \multicolumn{2}{c}{\textbf{WER (\%)}}  \\ 
            \cmidrule{4-5}
                                                &           &       & \textbf{Dev} & \textbf{Test} \\ 
            % \hline\hline
            % \multicolumn{5}{l}{\it Prior studies} \\ 
            \midrule
            Vanilla Whisper                     & -         & -         & 54.2  & 54.3              \\
            Separation + Whisper           
                                                            & -         & -     & \textbf{18.5}  & \textbf{15.3}              \\
            WavLM Base~\cite{zhang2023weakly}   & CLN       & 95.34 M   & -     & 27.5              \\
            TS-HuBERT~\cite{zhang2023weakly}    & CLN       & 105.18 M  & -     & 24.8              \\ 
            % \hline\hline
            % \multicolumn{5}{l}{\it Baseline TS-Whisper models} \\ 
            \midrule
            \multirow{4}{*}{TSE-Whisper (Full)}  & Add       & 762.98 M  & 26.6  & 28.3              \\
                                                & Cat       & 763.64 M  & 29.1  & 30.2              \\ 
                                                & FiLM      & 762.85 M  & 27.5  & 28.8              \\ 
                                                & CLN       & 763.38 M  & 27.3  & 28.8              \\ 
            % \hline\hline
            % \multicolumn{5}{l}{\it LoRA based fine-tuning} \\ 
            \midrule
            \multirow{4}{*}{TSE-Whisper (LoRA) } & Add       & 19.53 M   & 25.0  & 27.9              \\
                                                & Cat       & 20.19 M   & 28.7  & 31.2              \\
                                                & FiLM      & 19.40 M   & 24.9  & 25.6              \\
                                                & CLN       & 20.98 M   & 24.8  & 27.8     \\
            \bottomrule
        \end{tabular}
    }
    \label{tab:baseline_adapter_revise}
\end{table}
Table~\ref{tab:baseline_adapter_revise} presents the comparison results of different TSE modules.
For more details on these TSE modules, refer to Section~\ref{sec:tse_whisper}.
To evaluate performance, we also include results from prior studies utilizing the same TSE based target-speaker adaptation\footnote{Huang et al.~\cite{huang2023adapting} use the clean sets of Libri2Mix, so we exclude their results.}.
It can be seen that in the absence of target speaker clues, the vanilla Whisper model encounters challenges in the TS-ASR task, consistently predicting identical results for different speakers.
By integrating various adaptation methods, the TSE-Whisper models can effectively distinguish target speech, leading to a significant improvement from WERs of 54.2\% and 54.3\% to 26.6\% and 28.3\% on the Dev and Test sets, respectively.
However, there remains a substantial gap compared to the pipeline system, which first processes the mixture speech using a pre-trained target-speaker separation model\footnote{https://github.com/espnet/espnet/tree/master/egs2/librimix/tse1} and then performs single-speaker ASR with Whisper.
The gap may caused by the larger training data used in the separation model and the capacity limitations of our TSE-Whisper model.

Among the different types of the TSE module, the ``Add" approach demonstrates superior performance for fully fine-tuned models, while the ``FiLM" method proves most effective for LoRA fine-tuned models.
It is noteworthy that, even with approximately 3\% trainable parameters, LoRA fine-tuned models outperform fully fine-fined models across all adaptation methods.
This is likely because the TSE adaptation is susceptible to over-fitting during fully fine-tuning a large-scale speech foundation model, as they use fixed pre-extracted speaker embeddings as target-speaker enrollments, overlooking the variability introduced by interfering speakers.
% When compared with prior studies, our TS-Whisper models show inferior results despite having $7\times$ larger trainable parameters.
When compared with prior studies, our TSE-Whisper (LoRA) models show comparable results but still fall short of the best TS-HuBERT model.
This discrepancy can be attributed to the fact that both the WavLM base and TS-HuBERT models are pre-trained using multi-speaker mixed speech to learn a denoising capability, while the Whisper model lacks.

\subsection{Results of the SQ-Whisper model} \label{subsec:sq_whisper_results}

\begin{table}[tb]
    \centering
    \caption{Word error rates (WERs \%) of our proposed SQ-Whisper models on the noisy Libri2Mix sets. ``\#Param" refers to the number of trainable parameters. ``Full" means fully fine-tuning all parameters, while ``LoRA" means LoRA fine-tuning. ``SP" is the speed perturbation for data augmentation.}
    \resizebox{1.0\linewidth}{!}{
        \begin{tabular}{lcccc} 
            \toprule
            \multirow{2}{*}{\textbf{Model}} & \multirow{2}{*}{\textbf{Adapt. Method}} & \multirow{2}{*}{\textbf{\#Param}} & \multicolumn{2}{c}{\textbf{WER (\%)}}   \\ 
            \cmidrule{4-5}
                                                    &                   &           & \textbf{Dev} & \textbf{Test} \\ 
            % \hline\hline
            % \multicolumn{5}{l}{\it Prior studies}   \\
            \midrule
            Separation + Whisper                    & -         & -         & 18.5  & 15.3 \\         
            PIT-Transformer~\cite{zhang2023weakly}  & -                 & 32 M      & -     & 50.1  \\
            ~ + Train-$360$ \& SP \& LM             & -                 & 32 M      & -     & 23.5  \\
            WavLM Base~\cite{zhang2023weakly}       & CLN               & 95.34 M   & -     & 27.5  \\
            TS-HuBERT~\cite{zhang2023weakly}        & CLN               & 105.18 M  & -     & 24.8  \\
            PT-Whisper~\cite{ma2024extending}       & Prompt Tuning     & 51.82 M   & -     & 30.7  \\
            % x Whisper Large~\cite{ma2024extending}    & Prompt Tuning     & 107.11 M  & -     & 30.7  \\    
    
            % \hline\hline
            % \multicolumn{5}{l}{\it Baseline TS-Whisper models}   \\
            \midrule
            TSE-Whisper (Full)                       & Add              & 762.85 M  & 26.6  & 28.3  \\
            TSE-Whisper (LoRA)                       & FiLM              & 19.40  M  & 24.9  & 25.6  \\
    
            % \hline\hline
            % \multicolumn{5}{l}{\it Proposed TS-Whisper models}   \\
            \midrule
            SQ-Whisper (Full)                       & SQ-Former         & 793.64 M  & 20.2              & 20.1  \\
            ~ + Train-$360$ \& SP                   & SQ-Former         & 793.64 M  & \textbf{14.7}     & \textbf{14.6}  \\
            SQ-Whisper (LoRA)                       & SQ-Former         & 40.76 M   & 23.5              & 23.2  \\
            ~ + Train-$360$ \& SP                   & SQ-Former         & 40.76 M   & 16.8              & 16.0  \\
            SQ-Whisper (LoRA)                       & SQ-Former (1L)    & 26.19 M   & 24.0              & 23.6  \\
            \bottomrule
        \end{tabular}
    }
    \label{tab:sqformer_adapter_revise}
\end{table}

Table~\ref{tab:sqformer_adapter_revise} illustrates the WER results of our proposed SQ-Whisper models.
% Compared with the baseline models using FiLM adaptation, the SQ-Former adapted TS-Whisper models give a significant improvement on both fully and LoRA fine-funed conditions.
For the fully fine-tuned models, the SQ-Whisper gives a significant improvement over the baseline TSE-Whisper, reducing the WERs from 26.6\% and 28.3\% to 20.2\% and 20.1\% on the Dev and Test sets, respectively.
Compared with the previous best TS-HuBERT model, it also brings about 19\% relative WER improvement, which demonstrates its efficacy.
Moreover, our SQ-Whisper even surpasses the PIT-Transformer model trained with more data, reducing the WER from 23.5\% to 20.1\% on the Test set, yielding approximately 15\% relative WER improvement.
Furthermore, by augmenting the training data with an additional Train-$360$ set and the speed perturbation (SP) technique, we attain new state-of-the-art (SOTA) results, with WERs of 14.7\% and 14.6\% on the Dev and Test, respectively.
% Notably, our SQ-Whisper not only outperforms the pipeline system (14.6\% vs. 15.3\% on the Test set) but also eliminates the need for parallel separation training data, simplifies the training process, and reduces computational resources.
Notably, our SQ-Whisper also achieves better performance than the pipeline system (14.6\% vs. 15.3\% on the Test set) while being more efficient due to its end-to-end architecture and not requiring parallel separation training data.

Despite the impressive performance of our fully fine-tuned SQ-Whisper, it also introduces a large number of trainable parameters compared to prior studies ($7\times$ larger than the TS-HuBERT).
To ensure that the gains are not solely due to the increased parameters, we conduct experiments from two directions: using PEFT and reducing the number of SQ-Former layers to maintain a similar parameter budget.
% Therefore, we also explore parameter-efficient fine-tuning methods.
% In the case of LoRA fine-tuned TS-Whisper models, the SQ-Former adaptation continues to show superior results over the TS-HuBERT model with only half of the trainable parameters (50.19M vs. 105.18M), resulting in about 6.5\% relative WER reduction.
In the case of PEFT, we employ the LoRA technique, whose details are described in Section~\ref{subsec:peft}.
It can be observed that the LoRA fine-tuned SQ-Whisper continues to show superior results over the TS-HuBERT model with only half of the trainable parameters, reducing the WER from 24.8\% to 23.2\% on the Test set, giving about 6.5\% relative WER reduction.
Additionally, we fine-tune the model with a 1-layer SQ-Former to restrict the parameter size compared to the baseline TSE-Whisper (LoRA).
With almost similar parameters (21.60 M vs. 19.40 M), the SQ-Whisper still yields better results (23.6\% vs. 25.6\% on the Test set).

\subsection{Ablation study on the number of trainable queries} \label{subsec:num_queries_results}

\begin{table}[tbp]
\centering
% \caption{Ablation study on the impact of the quantity of learned queries. All hyper-parameters, except for the number of queries, remain constant across different models. The evaluation criterion utilized is word error rates (WERs).}
\caption{Word error rates (WERs \%) of our proposed SQ-Whisper with varying numbers of trainable queries on the noisy Libri2Mix sets.}
    \begin{tabular}{lccccccc} 
    \toprule
    \multirow{2}{*}{\textbf{Sets}} & \multicolumn{7}{c}{\textbf{Number of Queries ($L_q$)}}                 \\ 
    \cmidrule{2-8}
                          & 1    & 2    & 4    & 8    & 16   & 32   & 64            \\ 
    \midrule
    Dev                   & 23.7 & 22.5 & 23.3 & 21.5 & \textbf{20.2} & 20.8 & 20.5          \\
    Test                  & 23.9 & 22.0 & 22.0 & 21.1 & \textbf{20.1} & 20.8 & 20.8          \\
    \bottomrule
    \end{tabular}
    \label{tab:num_queries}
\end{table}

\begin{figure}[tbp]
    \centering
    \includegraphics[width=1.0\linewidth]{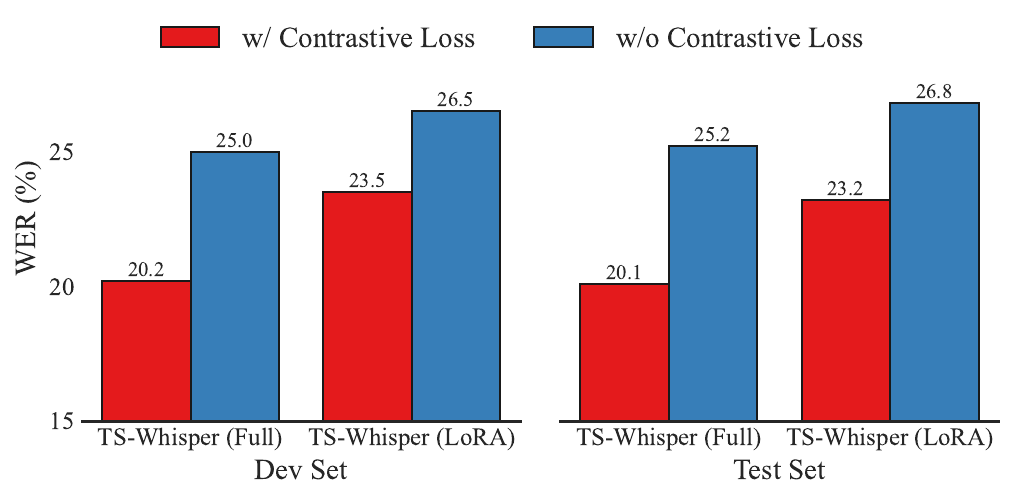}
    \caption{Word error rates (WERs \%) of our proposed SQ-Whisper with or without speaker contrastive loss.}
    \label{fig:cons_loss}
    % \vspace{-0.5cm}
\end{figure}

Here, we explore how the number of trainable queries $L_q$ affects model performance.
Specifically, we conduct experiments by fully fine-tuning the SQ-Whisper model, while varying the number of trainable queries within the range of \{1, 2, 4, 8, 16, 32, 64\}.
Table~\ref{tab:num_queries} summarizes the results.
Our analysis reveals a notable trend: as the quantity of trainable queries increases, there is a corresponding improvement of WERs, with a decrease from 23.7\% to 20.2\% on the Test set.
However, performance begins to deteriorate once the number exceeds 16.
This suggests that a small number of queries can effectively capture target speaker attributes, whereas an excessive quantity may introduce noise or lead to over-fitting.

\begin{figure}[tbp]
    \centering
    \subfloat[w/ Contrastive Loss]{\includegraphics[width=0.9\columnwidth]{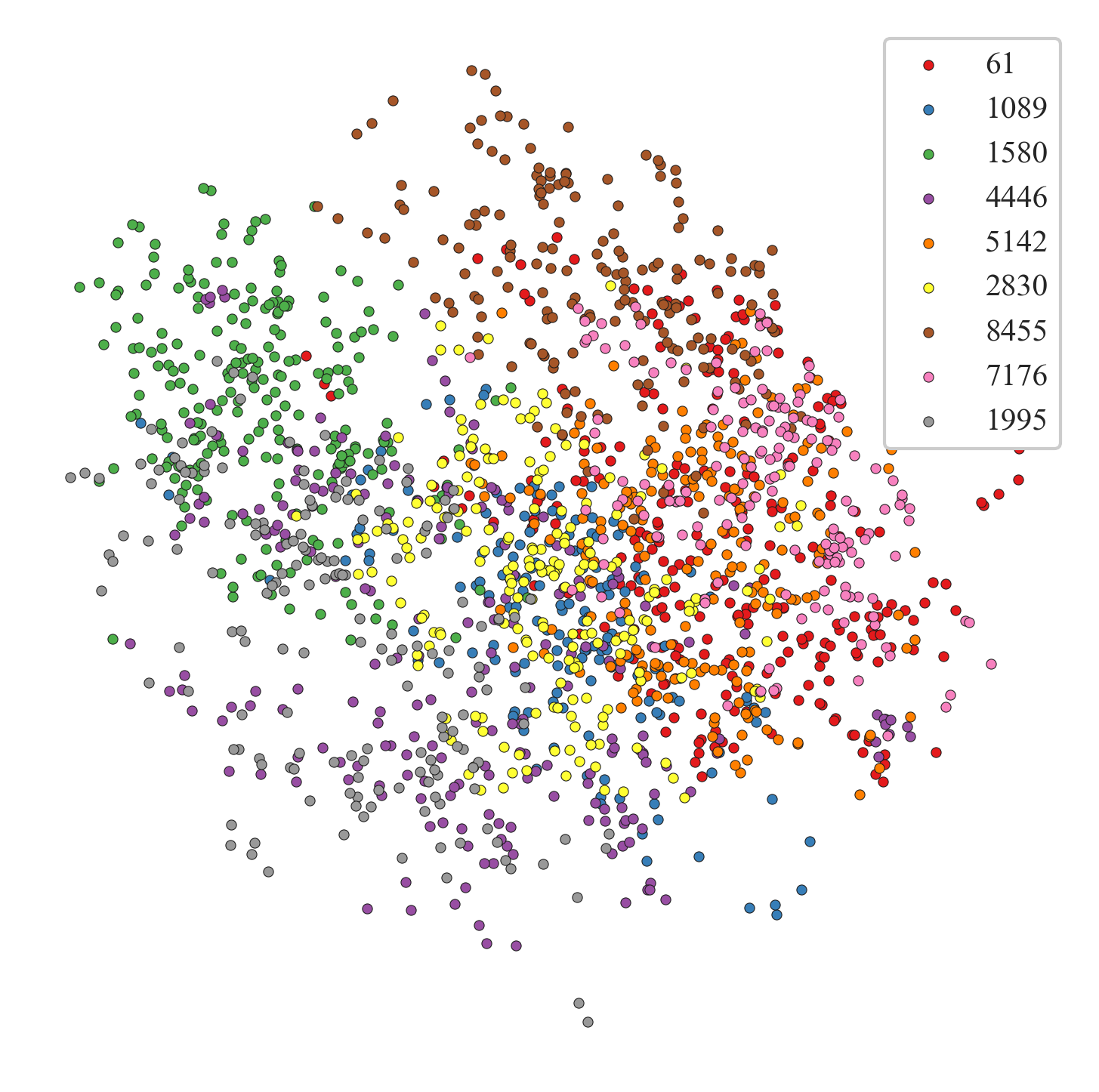}} \\
    \subfloat[w/o Contrastive Loss]{ \includegraphics[width=0.9\columnwidth]{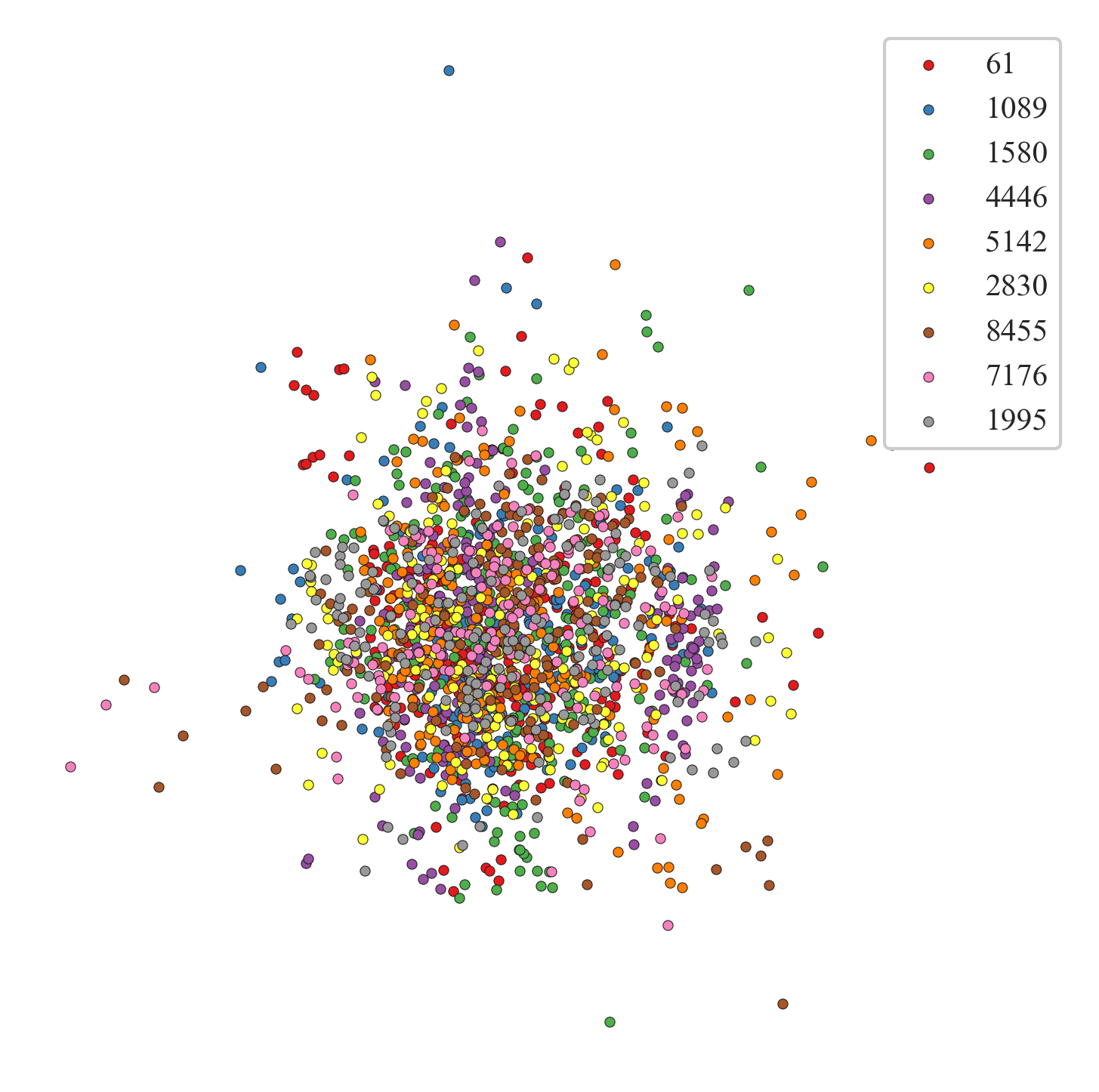}}
    \caption{T-SNE visualization of the learned speaker prompts with or without speaker contrastive loss. The legend refers to the speaker ID. A noticeable enhancement in distinctiveness is observed upon incorporating the speaker contrastive loss.}
    \label{fig:tsne}
\end{figure}

\subsection{Ablation study on the speaker contrastive loss} \label{subsec:con_loss_results}

% In contrast to the original Q-Former design, which aligns different modalities and learns visual representations most relevant to the text~\cite{li2023blip}, our goal is to utilize the learned queries to retrieve the attributes of target speakers from a mixture speech input based on an enrollment.
% In other words, the learned queries perform like a search engine.
% When provided with a search term (i.e., an enrollment of the target speaker), they are capable of retrieving desired relevant results (i.e., the learned target speaker prompts) from a database (i.e., the mixture speech input).

To generate discriminative speaker prompts, we introduce an additional speaker contrastive loss $\mathcal{L}_{\text{Contrastive}}$, as described in Section~\ref{subsec:con_loss}.
In this context, we investigate the influence of this loss.
Initially, we compare the results of the SQ-Whisper with or without $\mathcal{L}_{\text{Contrastive}}$, as depicted in Fig.~\ref{fig:cons_loss}. 
It is evident that both fully fine-tuned and LoRA fine-tuned models suffer a dramatic performance degradation upon the removal of the contrastive loss, resulting in up to 5\% absolute WER increment on the Test set.
Next, we visualize the learned speaker prompts using t-SNE, as shown in Fig.~\ref{fig:tsne}.
Specifically, we randomly select 9 speakers from the Test set and extract the learned prompts for all utterances of these speakers.
The model used to extract speaker prompts is the fully fine-tuned SQ-Whisper model, with or without $\mathcal{L}_{\text{Contrastive}}$.
The visualizations show a noticeable improvement in the distinctiveness after incorporating the loss.
It is important to highlight that our speaker prompts differ from the traditional speaker embeddings, like i-vector~\cite{dehak2010front} and x-vector~\cite{snyder2018x}, since the prompts are derived based on both enrollments and multi-speaker mixture speech.
However, we are interested in exploring the potential to employ the speaker prompts in speaker-related tasks, such as speaker verification or speaker diarization, and we aim to address this in future research endeavors.

% \begin{figure}[tbp]
%     \centering
%     \includegraphics[width=0.8\linewidth]{Figures/tsne.pdf}
%     \caption{T-SNE visualization of the learned speaker prompts with or without speaker contrastive loss.
%     The legend refers to speaker ID. A noticeable enhancement in distinctiveness is observed upon incorporating the speaker contrastive loss.}
%     \label{fig:tsne}
%     % \vspace{-0.5cm}
% \end{figure}

% \usepackage{multirow}

\begin{table}[tbp]
    \centering
    \caption{Word error rates (WERs \%) of our proposed SQ-Whisper with different prompt schemes. The ``Enc. Prompt" or ``Dec. Prompt" refer to appending speaker prompts in the encoder or decoder, respectively.}
    \begin{tabular}{cccc}
        \toprule
        \multirow{2}{*}{\textbf{Enc. Prompt}} & \multirow{2}{*}{\textbf{Dec. Prompt}} & \multicolumn{2}{l}{\textbf{WER (\%)}}  \\
        \cmidrule{3-4}
                        &              & \textbf{Dev}           & \textbf{Test}              \\
        \midrule
        \ding{56}       &  \ding{56}   &  54.2                  &  54.3                               \\
        \ding{52}       &  \ding{56}   &  21.9                  &  21.8                                \\
        \ding{56}       &  \ding{52}   &  27.6                  &  27.9                                \\
        \ding{52}       &  \ding{52}   &  \textbf{20.2}         &  \textbf{20.1}                                \\
        \bottomrule
    \end{tabular}
    \label{tab:prompt_scheme}
\end{table}

\begin{table}[tbp]
    \centering
    % \captionsetup{width=5cm}
    % \topcaption{Word error rates (WERs \%) of our proposed SQ-Whisper under matched and mismatched enrollment conditions. The ``Mismatched" condition refers to a scenario where the enrollment speaker is not included in the mixture speech, which means a randomly selected target speaker is used.}
    \caption{Word error rates (WERs \%) of our proposed SQ-Whisper under matched and mismatched enrollment conditions.}
    % \caption{Word error rates (WERs \%) of our proposed SQ-Whisper under matched and mismatched enrollment conditions. The ``Mismatched" condition refers to a scenario where the enrollment speaker is not included in the mixture speech, which means a randomly selected target speaker is used.}
    % \caption{Word error rates (WERs \%) of our proposed SQ-Whisper under matched and mismatched enrollment conditions. The ``Matched" condition indicates that the target speaker is present in the mixture speech, while the ``Mismatched" condition refers to a scenario where the enrollment speaker is not included in the mixture speech.}
    \begin{tabular}{lcc}
        \toprule
        \multirow{2}{*}{\textbf{Condition}} & \multicolumn{2}{c}{\textbf{WER (\%)}}  \\
        \cmidrule{2-3}
                                            & \textbf{Dev}      & \textbf{Test}       \\
        \midrule
        Matched                  & \textbf{20.2}     & \textbf{20.1}                \\
        Mismatched               & 71.6              & 71.8                 \\
        \bottomrule
    \end{tabular}
    \label{tab:untarget_spk}
\end{table}

\subsection{Ablation study on the prompt scheme} \label{subsec:prompt_scheme_results}

We further explore the impact of various prompt schemes, as summarized in Table~\ref{tab:prompt_scheme}.
For the ``Enc. Prompt", the learned speaker prompts are appended before the mixture features, while the ``Dec. Prompt" scheme means adding the prompts between the embedding space of $\langle \text{Prev} \rangle$ and $\langle \text{SOT} \rangle$ tokens.
For more details on the prompt schemes, please refer to Section~\ref{subsec:prompt_scheme}.
When employing prompts solely in the encoder or decoder, we observe that the ``Enc. Prompt" yields better results compared to its decoder counterpart (21.9\%/21.8\% vs. 27.6\%/27.9\%).
This suggests that the encoder, which captures acoustic information, plays a crucial role in guiding the model to distinguish target speech.
Furthermore, both prompt schemes result in a significant improvement over the original Whisper model, which lacks target speaker prompts.
Moreover, the ``Enc. Prompt" and ``Dec. Prompt" schemes demonstrate complementary effects, as their combination leads to further performance improvement.
An interesting thing is that despite the use of different prompt methods, we arrive at a similar conclusion as presented in~\cite{ma2024extending}.

\begin{table}[tb]
    \centering
    \caption{Word error rates (WERs \%) of our proposed SQ-Whisper models on the WSJ0-2Mix sets. ``\#Param" refers to the number of trainable parameters. ``LoRA" means LoRA fine-tuning. ``SP" is the speed perturbation for data augmentation.}
    \begin{tabular}{lccc} 
        \toprule
        \textbf{Model}    & \textbf{Adapt. Method} & \textbf{\#Param}     & \textbf{Test}         \\ 
        % \hline\hline
        % \multicolumn{4}{l}{\it Prior studies}  \\ 
        \midrule
        DPRNN-ASR~\cite{shi2022train}           & -                      & -                    & 7.1                   \\
        PIT-ASR~\cite{chang2021exploration}     & -                      & -                    & 12.1                  \\
        WavLM Base~\cite{zhang2023weakly}       & CLN                    & 95.34 M              & 10.4                  \\
        TS-HuBERT~\cite{zhang2023weakly}        & CLN                    & 105.18 M             & 6.1                   \\ 
        % \hline\hline
        % \multicolumn{4}{l}{\it TS-Whisper models}                                            \\ 
        \midrule
        % TS-Whisper (Full)                       & FiLM                   & 762.85 M             & 10.4                  \\
        Vanilla Whisper                       & -                         & -             & 72.0                   \\
        TSE-Whisper (LoRA)                       & FiLM                   & 19.40 M              & 7.6                   \\ 
        SQ-Whisper (LoRA)                       & SQ-Former              & 40.76 M              & 5.5                   \\
        ~ + SP                                  & SQ-Former              & 40.76 M              & \textbf{4.4}                   \\
        % \hline\hline
        % \multicolumn{4}{l}{\it Porposed TS-Whisper models}                                            \\ 
        % \midrule
        % TS-Whisper (Full)                       & SQ-Former              &                      &                       \\
        % ~ + SP                                  &    &  &   \\
        % TS-Whisper (LoRA)                       & SQ-Former              &                      &                       \\
        % ~ + SP                                  &                        &                      &                       \\
        \bottomrule
    \end{tabular}
    \label{tab:wsj_result}
\end{table}

\begin{table}[tb]
    \centering
    \caption{Word error rates (WERs \%) of our proposed SQ-Whisper models on the AMI-SDM1 sets. ``\#Param" refers to the number of trainable parameters. ``LoRA" means LoRA fine-tuning. ``SP" is the speed perturbation for data augmentation.}
    \begin{tabular}{lccc} 
        \toprule
        \textbf{Model}    & \textbf{Adapt. Method} & \textbf{\#Param}     & \textbf{Test}         \\ 
        % \hline\hline
        % \multicolumn{4}{l}{\it Prior studies}  \\ 
        \midrule
        Single-Speaker Conformer~\cite{kanda2021large} & -                      & -                    & 25.8                  \\
        SOT-Conformer~\cite{kanda2021large}   & -                  & -                    & \textbf{21.2}         \\
        % SOT-Whisper~\cite{li2023adapting}     & -  & 781.20 M                    & 23.6                  \\
        SOT-Whisper~\cite{li2023adapting}     & Residual Adapter  & 25.2 M                    & 25.0                  \\
        % \hline\hline
        % \multicolumn{4}{l}{\it TS-Whisper models}                                            \\ 
        \midrule
        % TS-Whisper (Full)                       & FiLM                   & 762.85 M             & 10.4                  \\
        Vanilla Whisper                        & -                      & -                  & 51.1                  \\
        TSE-Whisper (LoRA)                      & FiLM                   & 19.40 M             & 23.6                   \\ 
        SQ-Whisper (LoRA)                       & SQ-Former              & 40.76 M             & 22.3                   \\
        ~ + SP                                  & SQ-Former              & 40.76 M             & 22.0           \\

        \bottomrule
    \end{tabular}
    \label{tab:ami_result}
\end{table}

\subsection{Effectiveness Analysis of the SQ-Former Module} \label{subsec:effective_sqformer}
In this part, we investigate the effectiveness of the proposed SQ-Former module from three perspectives: robustness, interpretability, and computational overhead.
% Here, we investigate the effectiveness of the proposed SQ-Former module.
Since the SQ-Former is designed to learn target-speaker prompts relevant to the enrollment, we evaluate its robustness by randomly selecting mismatched enrollments for each mixture speech during inference, that is, the target speaker is randomly selected and not present in the mixture.
Table~\ref{tab:untarget_spk} shows the WER results under matched and mismatched enrollment conditions.
With mismatched enrollments, the WER results increase significantly from 20.2\% and 20.1\% to 71.6\% and 71.8\% on the Dev and Test sets, respectively.
Upon analyzing the decoded results, we find that the model tends to produce the same output regardless of the mismatched enrollments, favoring longer speech segments or those with higher energy in the mixture.
This suggests the SQ-Former fails to generate relevant speaker features for mismatched enrollments and thus provides non-discriminative speaker prompts.
% These findings indicate that the SQ-Former effectively captures target-speaker features relevant to the enrollment; however, when provided with a mismatched enrollment, it fails to generate discriminative speaker prompts.

\begin{figure*}[tbp]
    \centering
    \includegraphics[width=0.8\textwidth]{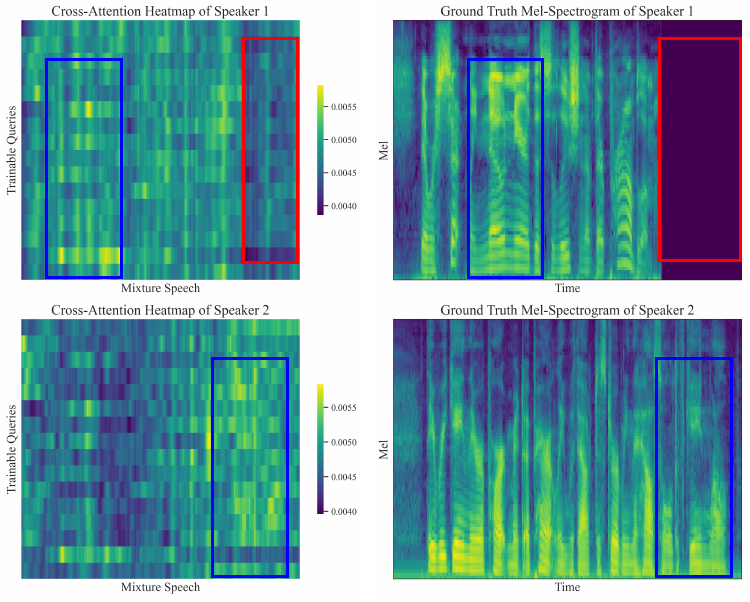}
    \caption{Visualization of the averaged cross-attention heatmap and ground truth Mel-spectrogram of each target speaker in the mixture speech. The selected mixture ID is 6930-76324-0001\_61-70970-0040.}
    \label{fig:attn_heatmap}
    % \vspace{-0.5cm}
\end{figure*}

To understand how the SQ-Former learns target-speaker prompts and its attention patterns, we visualize the averaged cross-attention scores in the SQ-Former alongside the ground truth Mel-spectrograms of each speaker in the mixture, as shown in Fig.~\ref{fig:attn_heatmap}. 
For ``Speaker 1" who speaks for a shorter duration, the trainable queries tend to attend more to the start of the mixture speech (blue box) while paying less attention to the end (red box), since the latter only contains speech of ``Speaker 2".
Conversely, for ``Speaker 2", the trainable queries focus more on the end of the mixture (blue box) because ``Speaker 1" has already finished speaking.

To quantify the additional computational cost introduced by the SQ-Former, we calculate the number of Floating Point Operations (FLOPs) for each model using a 10-second mixture speech input and a 10-second enrollment speech of the target speaker.
We utilize the open-source calflops\footnote{https://github.com/MrYxJ/calculate-flops.pytorch} toolkits for the FLOPs computation.
Specifically, the TSE-Whisper adds a modest computational overhead of 6.8\% compared to the original Whisper model, mainly due to the extra speaker embedding model and the FiLM adaptor.
Our SQ-Whisper introduces a computational overhead of 10.8\% compared to the original Whisper and 3.7\% compared to TSE-Whisper, attributable to the SQ-Former module.
We believe these increases are acceptable given the performance gains achieved.

\subsection{Results on the WSJ0-2Mix dataset} \label{subsec:wsj_results}
Table~\ref{tab:wsj_result} presents the WER results on the WSJ0-2Mix dataset.
For a quick evaluation, we only train models with LoRA fine-tuning.
While the TSE-Whisper gives a comparable result to prior studies, it still has a large gap with the powerful TS-HuBERT model (7.6\% vs. 6.1\%).
Our proposed SQ-Whisper achieves a best WER of 5.5\%, yielding about 10\% relative improvement compared to TS-HuBERT with only half of the trainable parameters.
By augmenting the training data with speed perturbation, we get further improvement, achieving a WER of 4.4\%.

\subsection{Results on the AMI dataset}
Table~\ref{tab:ami_result} shows the WER results on the real-world AMI meeting dataset.
Similar to the experiments on WSJ0-2Mix, we only train models with LoRA fine-tuning.
We can find that the SQ-Whisper model shows a consistent improvement over the TSE-Whisper model.
Compared with a strong single-speaker ASR model in~\cite{kanda2021large}, the SQ-Whisper gives up to 13\% WER improvement.
Besides, our SQ-Whisper model, which is pre-trained with 680k hours of single-speaker data, even achieves comparable results with the state-of-the-art model that is pre-trained with 900k hours of multi-speaker overlapping data (22.3\% vs. 21.2\%).

\section{Conclusion} \label{sec:conclusion}
In this study, we investigate the adaptation of speech foundation models for the recognition of multi-speaker overlapping speech, specifically targeting TS-ASR.
Using Whisper as the base foundation model, we first conduct a comprehensive exploration of its integration with established adaptation methods.
Subsequently, we introduce the innovative Speaker-Querying Whisper (SQ-Whisper) model, which effectively learns speaker prompts from a multi-speaker mixture speech and target-speaker enrollment speech.
The learned speaker prompts are used to steer the model toward extracting speaker-specific features and recognizing target-speaker transcriptions.
Experimental results on two simulation datasets, Libri2Mix and WSJ0-2Mix, as well as a real-world meeting dataset, AMI, demonstrate the effectiveness of our proposed method.
% achieving up to 15\% and 10\% relative WER reduction compared to the strong TS-Hubert model.
% Additionally, we also achieve a new state-of-the-art WER of 14.6\% on the Libri2Mix Test set after data augmentation. 
% Our findings highlight the potential of foundation models in complex multi-speaker scenarios, paving the way for more advanced ASR applications.
By leveraging the strengths of foundation models and inventive adaptation techniques, we contribute to the advancement of robust ASR systems capable of handling overlapping speech.
In future work, we plan to apply the SQ-Former to larger foundation models and extend the Whisper model to encompass multi-modality and multi-channel scenarios, thereby enhancing its applicability to more applications.

\bibliographystyle{IEEEtran}
\bibliography{refs}

\newpage

 

% \begin{IEEEbiography}[{\includegraphics[width=1in,height=1.25in,clip,keepaspectratio]{Figure/fig1.png}}]{Michael Shell}
% \end{IEEEbiography}



\vfill

\end{document}